 \documentclass[sigconf]{acmart}

\usepackage{xcolor}
\usepackage{framed}
\usepackage{tabularx}
\renewcommand\fbox{\fcolorbox{red}{white}}

\AtBeginDocument{%
  \providecommand\BibTeX{{%
    \normalfont B\kern-0.5em{\scshape i\kern-0.25em b}\kern-0.8em\TeX}}}

\copyrightyear{2023} 
\acmYear{2023} 
\setcopyright{acmlicensed}\acmConference[EASE '23]{Proceedings of the International Conference on Evaluation and Assessment in Software Engineering}{June 14--16, 2023}{Oulu, Finland}
\acmBooktitle{Proceedings of the International Conference on Evaluation and Assessment in Software Engineering (EASE '23), June 14--16, 2023, Oulu, Finland}
\acmPrice{15.00}
\acmDOI{10.1145/3593434.3593463}
\acmISBN{979-8-4007-0044-6/23/06}

\begin{document}








\title[Barriers for Social Inclusion in Software Engineering Communities]{Barriers for Social Inclusion in Online Software Engineering Communities - A Study of Offensive Language Use in Gitter Projects}



\author{Bastin Tony Roy Savarimuthu}
\affiliation{%
  \institution{University of Otago}
  \streetaddress{60 Clyde Street}
  \city{Dunedin}
  \country{New Zealand}}
\email{tony.savarimuthu@otago.ac.nz}

\author{Zoofishan Zareen}
\affiliation{%
  \institution{Independent Researcher}
  \city{Dubai}
  \country{United Arab Emirates}}
\email{zufishan.zareen5@gmail.com}

\author{Jithin Cheriyan}
\affiliation{%
  \institution{University of Otago}
  \streetaddress{60 Clyde Street}
  \city{Dunedin}
  \country{New Zealand}}
\email{jithin.cheriyan@postgrad.otago.ac.nz}

\author{Muhammad Yasir}
\affiliation{%
  \institution{Independent Researcher}
  \city{Wellington}
  \country{New Zealand}}
\email{yasir.muhammad1983@gmail.com}

\author{Matthias Galster}
\affiliation{%
  \institution{University of Canterbury}
  \streetaddress{Jack Erskine 314}
  \city{Christchurch}
  \country{New Zealand}}
\email{matthias.galster@canterbury.ac.nz}







\renewcommand{\shortauthors}{Savarimuthu, et al.}

\begin{abstract}
    Social inclusion is a fundamental feature of thriving societies. This paper first investigates barriers for social inclusion in online Software Engineering (SE) communities, by identifying a set of 11 attributes and organising them as a taxonomy. Second, by applying the taxonomy and analysing language used in the comments posted by members in 189 Gitter projects (with > 3 million comments), it presents the evidence for the social exclusion problem. It employs a keyword-based search approach for this purpose. Third, it presents a framework for improving social inclusion in SE communities. 
    
\end{abstract}

\ccsdesc[500]{Software and its engineering~Collaboration in software development}
\ccsdesc[500]{Software and its engineering~Programming teams}

\keywords{Social inclusion, exclusion, software engineering communities}


\maketitle

\section{Introduction}

Societies thrive if diverse individuals are able to engage with one another without any discrimination \cite{ostrom2008}. Prejudiced beliefs and discriminatory actions against individuals or groups lead to social exclusion in these communities \cite{perkins2013}, thus impacting diversity \cite{amin2019}.  Social exclusion is a phenomenon that occurs when someone is forcibly or voluntarily separated from groups with whom they interact with on a daily basis \cite{malik2019}. Social exclusion has been shown to lead to poor health, education, employment and economic outcomes for individuals and groups that are excluded \cite{worldbank2013}. Recently, there has been a surge in interest in investigating social issues in online SE communities including offensive language use \cite{cheriyan2021,miller2022,sultana2021,qiu2022}. The targets of offensive language are individuals and groups who possess a specific attribute (e.g., belonging to a specific race or religion) \cite{cheriyan2021}. Offensive remarks can make these community members feel discriminated and unwelcomed, paving way for the feeling of social exclusion \cite{bradshaw2004, appleton2014, ademiluyi2022}.


Researchers have noted that discrimination based on certain attributes creates unequal power relationships and these are underpinned by beliefs that groups with specific attributes are inherently superior, and these beliefs help maintain and legitimise a group-based hierarchy \cite{nz2020}. When individuals or minority groups see actions or hear comments that make them feel inferior they often feel excluded because of discrimination \cite{bradshaw2004, appleton2014, ademiluyi2022}. This social exclusion not only impacts an individual's present situation but denies opportunities for the future. While researchers have posited that digital inclusion of marginalized societies can improve their social inclusion \cite{ragnedda2022}, this work contends that the language use within online SE communities should be carefully managed, as improper use can trigger social exclusion. Thus, this work situated in the emerging domain of enhancing diversity and inclusion in SE communities \cite{bosu2019,serebrenik2020}, aims to investigate attributes that can trigger social exclusion in SE communities, and also quantify the prevalence of attribute-targeting problem (where individuals are targeted based on the attribute\footnote{An attribute is a specific characteristic of a person such as race and religion.} they possess) in these communities. Towards this end, this paper poses the following three research questions. 

\textbf{RQ1. What attributes when targeted in communication can offend individuals thus hampering social inclusion?}

\textbf{RQ2. What is the prevalence of socially-exclusionary messaging in Gitter, a popular online platform used by software engineering projects?} 

\textbf{RQ3. How can we start addressing the problem of socially-exclusionary messaging in software engineering communities?}




Through answering RQ1, this paper proposes a taxonomy of social exclusion attributes for systematically thinking about barriers for social inclusion in SE communities, based on data from Stack Overflow's rude and offensive comments. Also, using the taxonomy, it identifies the volume of comments on 189 projects hosted on the Gitter platform  that have the potential to trigger social exclusion feelings (thus answering RQ2). Finally, by answering RQ3, it sets an agenda for future work to create socially-inclusive SE communities. 

\section{Social Inclusion in SE communities}

Social inclusion is \emph{``the process of improving the ability, opportunity, and dignity of people,
disadvantaged on the basis of their identity, to take part in society''} \cite{worldbank2013}. New Zealand's social inclusion report commissioned as an immediate aftermath of the Christchurch shootings \cite{nz2020} defines this as \emph{``having an equitable opportunity to participate, by choice''}. We adapt these two definitions \cite{worldbank2013, nz2020} that stem from real-world, physically-based settings to suit the online context, by defining,  social inclusion in online communities (aka \emph{digital social inclusion}), as \textit{having an equitable opportunity to participate, by choice, in online activities of a community, without being disadvantaged on the basis of their identity and other factors (e.g., preferences)}.  Barriers for such equitable participation are prejudiced attitudes and discriminatory behaviours within a community that arise based on identity-based characteristics such as a person’s age, culture, beliefs, abilities (e.g., language use), gender identity, sexual orientation, appearance, family background, and income \cite{worldbank2013}. 

 SE communities, a subset of online communities, are of interest in this paper. Online SE communities have become a commonplace for different SE stakeholders to share existing knowledge and create new knowledge. These communities operate on well-known platforms such as Stack Overflow, GitHub, Gitter and Slack.  It has been observed that SE communities such as Stack Overflow have become an unwelcoming place. Jay Hanlon, the executive vice president of Stack Overflow noted that \emph{``Too many people experience Stack Overflow as a hostile or elitist place, especially newer coders, women, people of color, and others in marginalized groups''} \cite{hanlon2018}. While this description points to factors triggering social exclusion, prior work has not systematically identified these factors. This work is our initial attempt to bridge this gap by proposing a taxonomy of attributes that contribute towards social exclusion, which is presented in the next section. 
 

\section{A taxonomy of social exclusion attributes in SE communities}

\begin{figure} [htb]
  \includegraphics[width=0.5\textwidth]{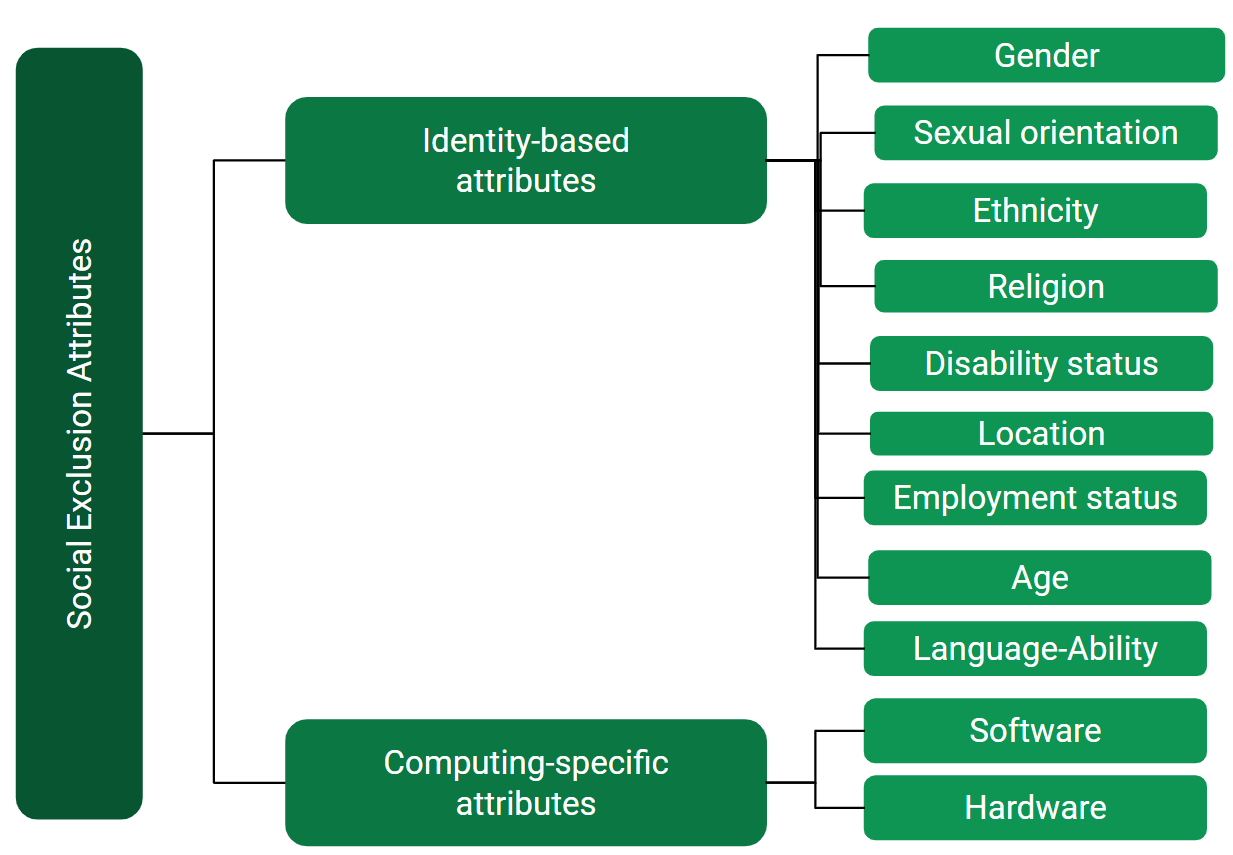}
  \caption{Social exclusion taxonomy for Software Engineering}
  \label{fig:taxonomy}
\end{figure}

The first goal of this work is to present a taxonomy of attributes of an individual or a group, that when targeted (e.g., age) can contribute towards social exclusion, thus answering RQ1. We have created a hierarchical taxonomy \cite{usman2017} containing 11 attributes (leaf nodes as shown on the right of Figure \ref{fig:taxonomy}). The taxonomy shows two types of social exclusion attributes - an individual's \emph{identity-based attributes} (e.g., race) and \emph{computing-specific attributes} (attributes that are specific to computing that get targeted which are software and hardware).

The taxonomy was constructed based on a combination of top-down and bottom-up approaches, two commonly used approaches for building taxonomies \cite{capiluppi2020}. These two approaches are also known as enumerated (fixed) and  faceted (emergent) approaches (cf. \cite{usman2017}). Following the top-down approach, we used attributes that were identified based on existing literature on social exclusion in real-life societies \cite{nz2020,worldbank2013}. There were seven attributes identified using this approach (Gender, Sexual orientation, Ethnicity, Religion, Disability status, Location and Employment status). Then, using a bottom-up approach, we identified four  additional attributes based on themes that emerged  (Age, Language-ability, and the two computing-specific attributes) by analysing  200 randomly selected comments from the rude and offensive Stack Overflow dataset \cite{woah2020dataset}. These were identified by the first author and verified by the third author (100\% consensus).

\subsection{Identity-based attributes}
Below, we describe identity-based attributes that can trigger the feelings of social exclusion when a person that matches the identity is targeted using socially-exclusionary messages. The term “exclusionary message” is used in this paper for comments containing wording that discriminates individuals based on one or more attributes. Below, for each attribute, we provide an example comment posted on Stack Overflow which was tagged as rude or offensive by moderators which were subsequently removed \cite{sooffensive,woah2020dataset}.


\begin{itemize}
\item {\verb|Gender|}: This represents the biological characteristics of a person (e.g. male or female) or the internal sense of self that a person identifies with (e.g. a man identifying as a woman). Sample comment: \textit{\textcolor{blue}{Please explain to me the problem - it works for me. What is your fucking problem? You want something to work when you refuse to install the perquisites and you bitch about it like a little girl when it does not work?}}







\item {\verb|Sexual orientation|}: This attribute represents the identity of an individual based on their sexual preference such as gay, lesbian, bisexual, transgender, queer or intersex. 




Sample comment: \textit{\textcolor{blue}{Fuck you for down voting my question and i don't need your gay ass fucking help now because i already configured it.}}


\item {\verb|Ethnicity|}: This attribute identifies the ethnic group an individual belongs to. Examples include Indian, American, and African.  Sample comment: \textit{\textcolor{blue}{Fuck the Indian who down voted the question.}}




\item {\verb|Religion|}: This attribute identifies the religion one belongs to - e.g., Islam, Christianity and Hinduism. Sample comment: 

\textit{\textcolor{blue}{hello, do you want to tell me that in all this stupid shit internet there is not a single solution to this ridiculous simple problem of adding fucking products to this little pedophile christian script.}}



\item {\verb|Disability|}: This attribute identifies the physical or mental inability that impedes a person from fully participating in the society (e.g., wheelchair bound and dyslexia). Sample comment: \textit{\textcolor{blue}{@username if you actually read the question and looked at the picture, you wouldn't sound like a retard.}}





\item {\verb|Location|}: This attribute identifies a specific location a person is from (e.g., Africa and Pakistan). 

Sample comment: \textit{\textcolor{blue}{... go home to pakistan or india or whereever you dirty ass comes from. You should be sweeping our streets not lecturing me on sites like this...}}



\item {\verb|Employment status|}: This attribute identifies whether someone is employed or not, and if they work, whether they are a part-timer. Sample comment: \textit{\textcolor{blue}{Go find a hobby. You seem like the type who is on SO [Stack Overflow] way too much. May be get a job.}}



\item {\verb|Age|}: This attribute identifies the age group an individual belongs to (e.g., young vs. old, and specific age-groups such as teenagers). Sample comment: \textit{\textcolor{blue}{@username aged **60** years old person, answering questions. fuck off ...}}



\item{Language-ability}: This relates to the ability to use a particular language such as English during interactions. 
Sample comment: \textit{\textcolor{blue}{I feel like I'm not the only one who has no fucking clue what you want. Please ask your question in proper English and provide more detail ...}}





\end{itemize}


\subsection{Computing-specific attributes}
This is divided into two main categories: Software and Hardware as shown below.

\begin{itemize}
\item{Software}: This relates to derogatory comments made about specific software, sometimes, in comparison with competitors (Mac OS vs. Windows, Chrome vs. Firefox, and Java vs. Python). Also, these comments may relate to technical knowledge and skills of individuals, and also software development processes. Sample comment: \textit{\textcolor{blue}
{Where the fucking hell is this Label Letter Space on Xcode?}}

\item {Hardware}: This relates to derogatory comments made on specific hardware. Sample Comment: \textit{\textcolor{blue}
{``Have to buy a Mac (not necessarily a bad thing''. No, it's bullshit. Having to pay a yearly(!) fee to publish apps, and have them be rejected by Apple for an arbitrary reason, is bullshit. I find it amazing that so many developers are blinded by \$\$\$ to the point that they don't care about open platforms and open hardware.}}

\end{itemize}

\subsection{Validation of the taxonomy}
To validate the taxonomy, we conducted a small pilot study involving five professional software developers in New Zealand (3 males and 2 females) recruited using a convenience sampling approach. All had at least a bachelor's degree, and their work experience ranged from 2 to 15 years. The participants were given 11 offensive language sentences (one belonging to each attribute category). They were asked to report a) the level of offensiveness they felt after reading the offensive comments in two categories (identity-based and computing-specific factors) using a 4-point likert scale (very offensive, moderately offensive, mildly offensive and not offensive), and b) whether these offensive comments would have caused emotional distress among the victims and may  have lead them to not engage with the offending communities effectively (i.e., triggering social exclusion), which was a yes or no question. All the participants found the identity-targeting comments offensive - three felt strongly offended, one moderately offended and one mildly offended. For computing-specific offense categories the result was almost the same. Two were strongly offended, one moderately offended, and two mildly offended. Four out of five (80\%) agreed comments in these two categories may have caused emotional distress amongst the victims, and they may not have been able to engage with these communities effectively (i.e., these comments had the potential to trigger social exclusion). For the question whether they find any difference in the offensiveness level that they feel, when a person is targeted vs. a computing-specific attribute is targeted, three of those explicitly stated they find personal targeting more offensive. 




\noindent\fbox{%
    \parbox{\linewidth}{%
        \textbf{Key finding of RQ1}: The taxonomy presented and examples from Stack Overflow above indicate text posted in SE communities contain offensive messages that target specific socially-exclusionary attributes.
    }%
}

\section{Gitter study} \label{sec:gitter_study}
In order to demonstrate the practical utility of the taxonomy developed, we investigated the comments posted by Gitter developers on 189 projects, comprising a total of 3,014,999 comments, based on the data of Parra et al. \cite{parra2020}, thus answering RQ2. The Gitter data set was selected because it contained both offensive and non-offensive comments as opposed to the offensive comments data sets of Stack Overflow (e.g., \cite{sooffensive,woah2020dataset}).



We employed a keyword-based approach to identify socially-exclusionary messages posted in these communities. We followed a five-step process. In the \emph{first} step, we identified comments that contained offensive language phrases. We sourced the set of offensive keywords from various dictionaries\footnote{An offensive phrase list is \url{https://rdrr.io/cran/lexicon/man/profanity_alvarez.html}.} that are deemed to be offensive. Then, we identified all sentences in the Gitter data set that contained these offensive words, using regular expressions. There were 4,342 sentences in total. In the \emph{second} step, the identified sentences data set (i.e., 4,342 sentences) was partitioned into two sets, and two coders (second and fourth authors) read the comments in a set each to check whether these are offensive. They also completed a prior pilot task involving 100 comments to identify offensive comments, for which there was full consensus. Upon classification, we undertook a \emph{third} step to identify whether a comment  contained socially-exclusionary messaging targeting a particular attribute (e.g., an abusive comment targeting sexual orientation). To identify this, we used a keyword based approach where keywords representing different attributes were compiled and these were searched within the data set that have been deemed to be offensive. We created nine keyword lists, one pertaining to each identity-based attribute\footnote{\url{https://github.com/bastintonyroy/barriers_for_social_inclusion}}. Using regular expressions, we identified all offensive comments that contained the socially-exclusionary keywords. To identify offensive comments related to computing-specific categories (i.e., hardware and software), due to the varied nature of the names of hardware and software used in comments, two coders (second and fourth authors) read through 4,342 comments to identify whether these are offensive. At the end of this step, in total, we identified 605 comments that belonged to these 11 attribute categories.


 \emph{Fourth}, another round of validation was conducted where all these 605 comments that contained one or more attributes (out of 11) were further checked by two coders (first and third authors) based on a rubric developed (see footnote 3).  There was 98\% agreement between the coders. Full consensus was achieved after discussions with the final list containing 605 comments.  Representative comments for the 11 attributes from Gitter projects are shown in Table \ref{tab:1}. Finally, as the \textit{fifth} step, we created an RShiny app that makes these comments available to the public\footnote{\url{https://socialsoftware.shinyapps.io/Social_Exclusion_Category_Viewer/}}. 



\begin{table*}[t]
  \centering
\begin{tabular}{|p{2.5cm}|p{13.5cm}|}
  \hline
 \textbf{Attribute} & \textbf{Comments on Gitter} \\
 \hline
 Gender & \textcolor{blue}{\emph{so... is the boob operator seriously being considered?}}
\\
 \hline
 Sexual orientation  & \textcolor{blue}{\emph{... you should just make it work first, you can make it all gay later, css rainbows and whatnot}}
\\
  \hline
 Ethnicity  & \textcolor{blue}{\emph{if you don't have nazi colleagues who review your code :)}}  \\
  \hline
 Religion  & \textcolor{blue}{\emph{jesus fucking christ}}  \\
  \hline
 Disability status  & \textcolor{blue}{\emph{i feel like a retard lol. i usually pick thing up after just 2-3 times but with margin, padding, border and... whatever... i have been on it like 50 times and still cannot remember how it work lol}}  \\
  \hline
 Location  & \textcolor{blue}{\emph{fucking china attacked github}}   \\
  \hline
 Employment status  & \textcolor{blue}{\emph{i've been off technical work for about 5 years. i find my internal geek unfulfilled with a  job that is interesting, but a bit too much bureaucratic for my taste. i need something more. so that is my challenge: make a meteor app that does my job for me, learn a shitload, understand the new way and then try and do some shit with the occulous rift, cause the web is going bye bye.}}   \\
  \hline
 Age  & \textcolor{blue}{\emph{and then i feel old as shit and stop myself}}   \\
  \hline
 Language-ability  & \textcolor{blue}{\emph{I can't decide if i'm more pedantic or more non-english speaker. do i give a fuck about this? *should* i give a fuck about this?
}}  \\
  \hline
 Software  & \textcolor{blue}{\emph{Is that why js and html5 and transpilers  have almost reproduced silverlight except slower and with shitty debugging?
}} \\
  \hline
Hardware  & \textcolor{blue}{\emph{what the fuck do you mean you want to update my windows? sick bastard pc}}   \\
\hline
\end{tabular}
\caption{socially-exclusionary comments on Gitter}
  \label{tab:1}
\end{table*}



\begin{figure}[htb]
  \includegraphics[width=0.5\textwidth]{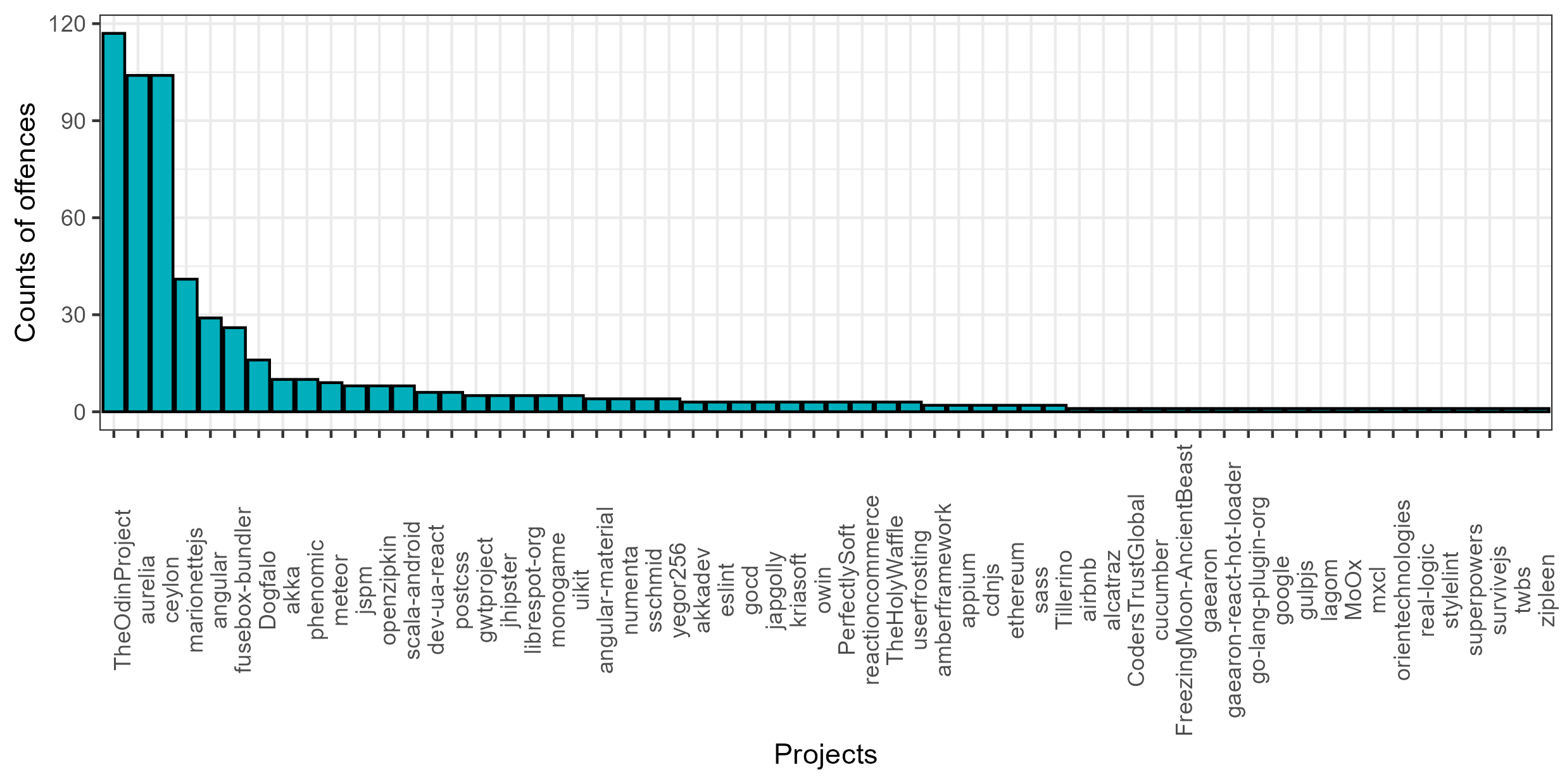}
  \caption{Offences per project}
  \label{fig:offences_per_project}
\end{figure}

We analysed the 605 comments that contained one or more social exclusion attributes. 
Figure \ref{fig:offences_per_project} shows the total counts of exclusionary messages in 189 projects. We observed that 60 out of 189 projects (32\%) contained socially-exclusionary comments. The top-3 projects had more than 100 such exclusionary comments. The top-9 projects had >=10 socially-exclusionary comments. 

\begin{figure}[ht]
  \includegraphics[width=0.5\textwidth]{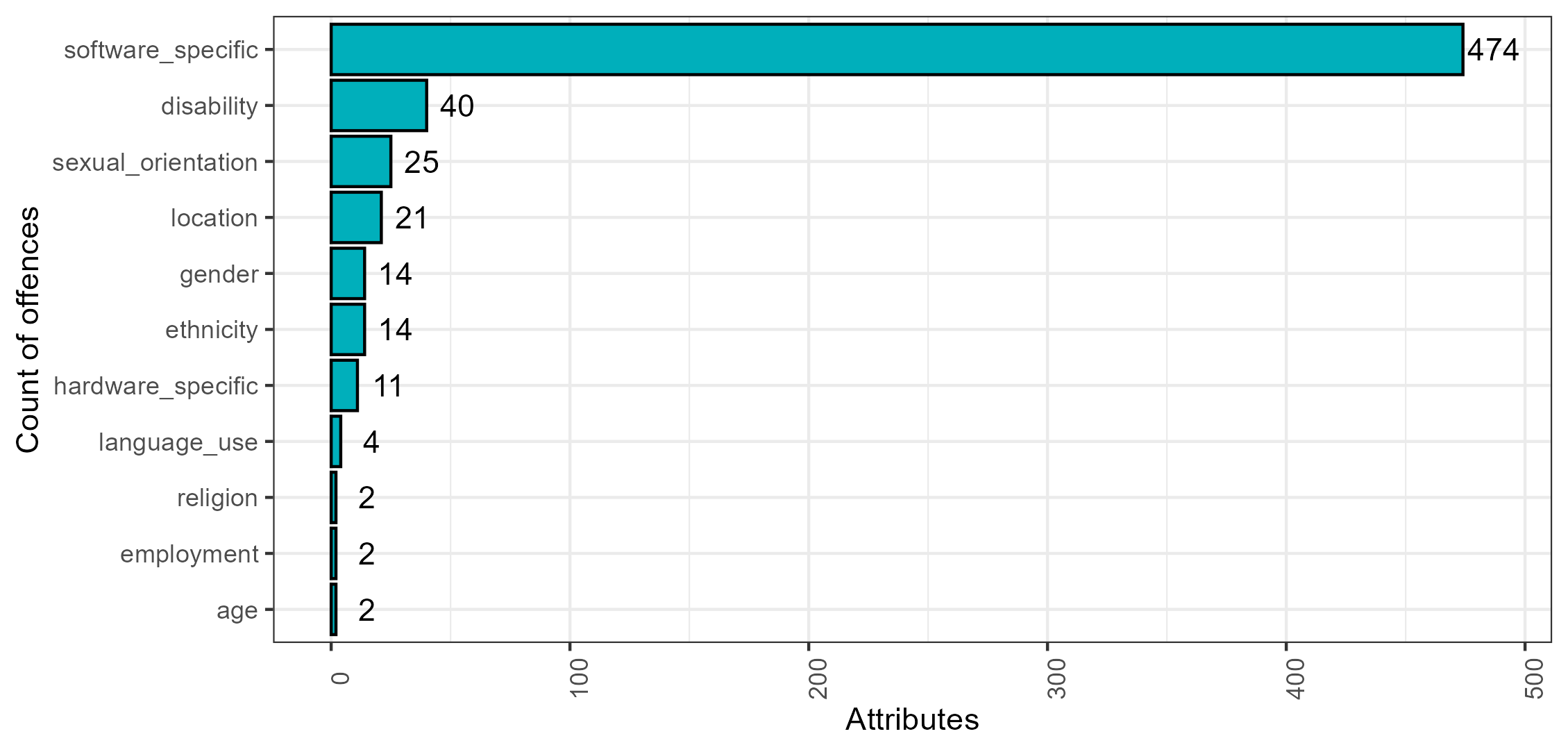}
  \caption{Offences count per attribute across all projects}
  \label{fig:total_offences}
\end{figure}

Figure \ref{fig:total_offences}  shows the counts of socially-exclusionary comments for all the attributes. Our results show that, in fact, all the 11 attributes were targeted within this data set. The top three attributes that were targeted are: software (474), disability (40) and sexual orientation (25). The attributes religion, employment and age were targeted the least. 

\begin{figure}[htb]
  \includegraphics[width=0.47\textwidth]{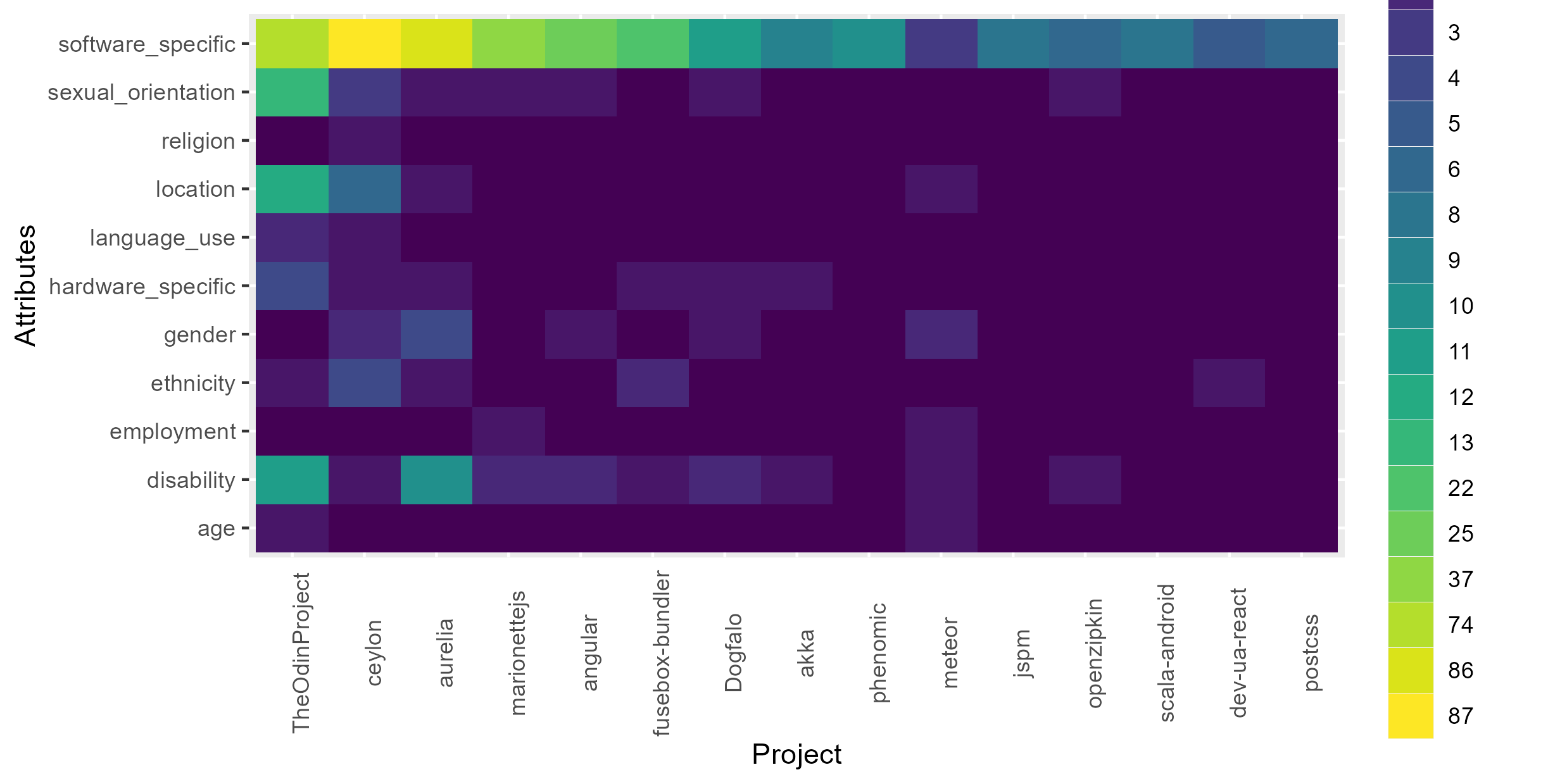}
  \caption{Offences by type (attribute) across top 15 projects}
  \label{fig:offence_by_type}
\end{figure}

Figure \ref{fig:offence_by_type} is a heatmap that shows the counts of different socially-exclusionary attributes present (y-axis) for the top-15 projects (x-axis). The projects on the x-axis are ordered based on the descending order of the number of offensive comments. The three projects that had the most offensive comments (The Odin Project, Ceylon and Aurelia), had comments that belonged to 7, 9 and 8 attribute categories (out of 11) respectively.  On the other hand the project ranked 15th (postcss) had comments targeting just one attribute.

\noindent\fbox{%
    \parbox{\linewidth}{%
        \textbf{Key finding of RQ2}: The results from Gitter projects confirm the relevance of the presented taxonomy, and its utility in identifying socially-exclusionary messages.
    }%
}

\section{Towards Addressing socially-exclusionary messaging}

The need for inclusivity in real-life societies has been acknowledged by governments and NGOs alike \cite{nz2020,worldbank2013}, needing long-term solutions. This issue has also been acknowledged in the SE research community, and as a response, call for action and a set of recommendations have been made \cite{serebrenik2020}. However, the efficacy of recommendations is not known yet as adoption and eventual evaluation require a protracted period of time. So, to bootstrap the process of finding a solution to address the issue of social exclusion in SE communities, we asked the same five developers involved in the evaluation of comments about how they would have felt if they were the recipients (i.e., targets) of socially-exclusionary comments and what change they would like to see in these communities. They all agreed on two aspects (100\% agreement): 1) they wished these offensive comments are not allowed in the first place and 2) they also hoped that there were mechanisms that paraphrased offensive comments to make them non-offensive. These point towards the need for developing a framework which can achieve these goals.

We propose a framework with four stages called DARE (Detect, Classify, Explain and Reduce) that can be used to manage and prevent such exclusionary messages. Below, we outline these stages. 

\begin{figure}[htb]
  \includegraphics[scale=0.26]{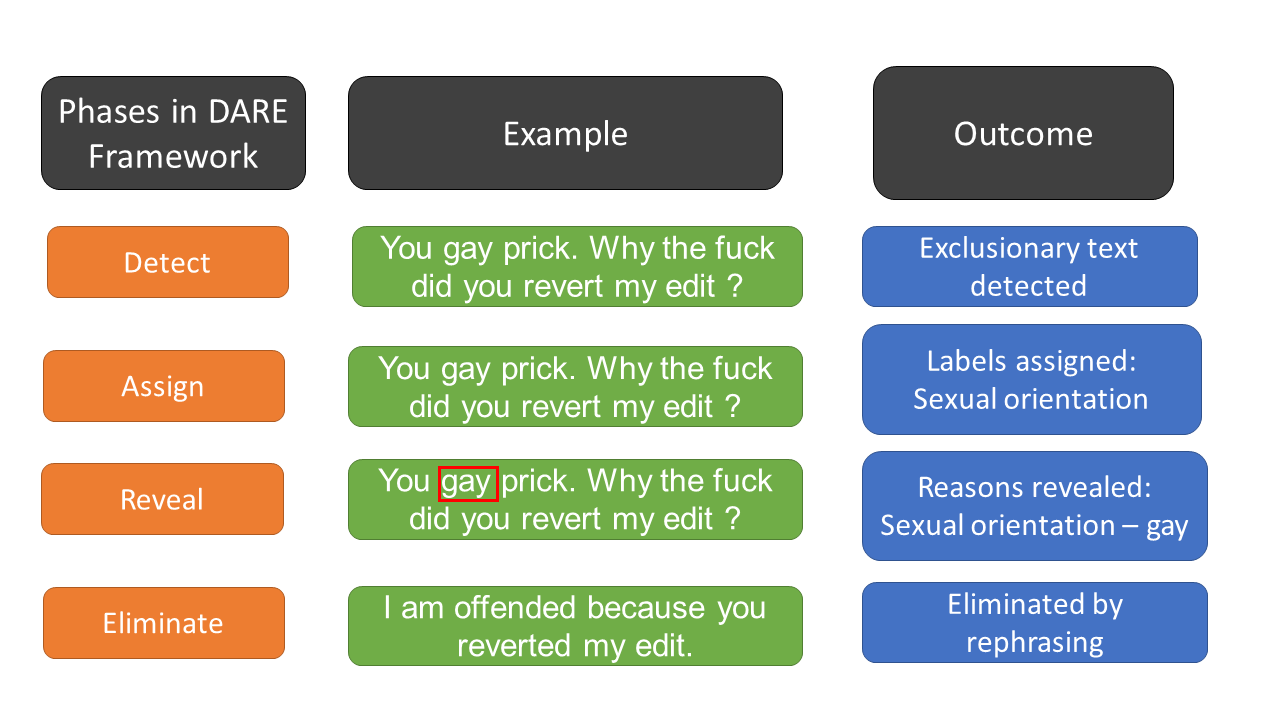}
  \caption{DARE framework for countering social exclusion in SE communities}
  \label{fig:dare_framework}
\end{figure}

The \emph{first} stage is to detect the presence of such exclusionary text (Detect stage) in SE platforms. Upon detection, the exclusionary text needs to be assigned label(s) in the \emph{Assign} stage. These labels are based on the 11 attributes identified in the taxonomy. The text can have multiple labels. This labelling approach can be useful to raise awareness about the offensive nature of the text to the poster. This is enabled in the \emph{third} stage (Reveal stage). In this stage, the reason for an offense is revealed to posters through highlighting the phrases in the text (e.g., by underlining or placing a box around) that are considered to be exclusionary. Thus, this stage offers explanation for a text being considered to be exclusionary. The final stage is the \emph{Eliminate} stage. In this stage, the comment poster can be provided with an opportunity to modify the offensive statement or an automatic rephrasing tool can be used to eliminate/eradicate aspects of the text that are perceived to be exclusionary. Figure \ref{fig:dare_framework} shows using an example, how this approach can be implemented. The stages are shown in the first column. The second column contains the example considered. The third column shows the outcome from the stages. The immediate action point in this line of work is to employ ML approaches to classify attributes and develop the full pipeline of the DARE framework.

\noindent\fbox{%
    \parbox{\linewidth}{%
        \textbf{Key outcome from RQ3}: In order to effectively manage socially-exclusionary messaging, a concrete step forward will be to implement computational-based frameworks such as DARE. 
    }%
}

\section{Discussion} \label{sec:related_work}

\textbf{Related work and contributions: } We were motivated by the emerging interest amongst SE researchers in investigating toxicity and offensive language in developer communications \cite{raman2020, miller2022,cheriyan2017, cheriyan2021, ferreira2021, ferreira2022,  qiu2022, sarker2022}. Prior works in this arena have considered developer exchanges in the text format from various sources such as GitHub issues \cite{miller2022}, Stack Overflow comments \cite{cheriyan2021}, Gitter discussions \cite{parra2020} and code review text \cite{qiu2022}. However, the focus has been mainly on identifying whether the communication text is offensive/toxic (e.g., \cite{raman2020, ferreira2022}), and not on presenting the reason behind the offense. Said that, there has been some attempt in identifying the reason for the text being considered to be offensive (e.g. use of racial, swearing and misogyny terms \cite{cheriyan2021, sultana2021}), and these works are directly relevant to our work here. In the same vein, the work of Miller et al. \cite{miller2022} has identified reasons for toxic comments on GitHub such as failed tool/code, technical disagreements, politics/ideologies and past interactions. However, these directly relevant works (\cite{cheriyan2021, sultana2021,miller2022}) do not view toxic or offensive language use from a social exclusion view point. Thus, the taxonomy presented in this work (as a response to RQ1) adds to the repertoire of work in the domain of diversity and inclusion. Also, our work here demonstrates the utility of the taxonomy by unearthing comments that belong to all the 11 socially-exclusionary attributes (thus answering RQ2). Our results show that some projects generate more socially-exclusionary messages, targeting multiple attributes than others, with some projects containing comments targeting 9 out 11 socially-exclusionary attributes. This shows there is a need for educating communities using norm and value-based arguments, which are fundamentally
about changing societies in a way that allows everyone
(independent of their attributes and preferences) to realise their potential. Specific recommendations made for real-life, physically-based societies \cite{nz2020}, can be applied to online SE communities to become more inclusive. These include making inclusive practices more explicit as a part of the Code of Conduct (CoC) to raise awareness, and when the norms are violated, the users should be reminded about CoC. Applying appropriate sanctions for norm violations should also be considered. An immediate first step to address the social inclusion issue (RQ3) is to adopt ML approaches in implementing the DARE framework. For example, ChatGPT API\footnote{https://openai.com/blog/introducing-chatgpt-and-whisper-apis}, can be trialled. 

\textbf{Future work:} While there have been strands of work that focus on specific attributes: e.g., gender issues  \cite{bosu2019,sultana2021}, there is scope for more work on other attributes such as racism, sexual orientation, and language-ability in SE. For example, prior work has noted that there are no explicit traces of racism \cite{rodriguez2021}, but our work here and others \cite{cheriyan2021} provide some contrary evidence. This warrants further investigation given the discussions on this topic in online forums\footnote{\url{https://www.reddit.com/r/cscareerquestions/comments/2ve8cs/racism_in_software_development/}}. A recent systematic literature review on diversity that focused on four attributes (age, race, nationality and disability), also notes that  perceived diversity aspects of SE participants such as their race, age, and disability
need to be examined more in SE research \cite{rodriguez2021}. 

\textbf{Threats to validity: } There are some known threats to the validity of our current work. While this work establishes the presence of 11 attributes in the data sets considered (Stack Overflow and Gitter), it is possible that there are more categories which have not been identified due to their absence in the comments considered in this work. This is a threat to \emph{construct validity}.  A threat to \emph{internal validity} is that we only considered text that contained specific keywords that signal offensiveness. Thus, the results above are conservative estimates of socially-exclusionary messages. To strengthen the \emph{external validity} of our taxonomy, we need to conduct larger studies involving comments from multiple platforms (e.g., GitHub and Slack), and also ask participants to identify attributes in text that can trigger feelings of social exclusion. 





\section{Conclusion}
The paper presents a taxonomy of 11 attributes that can trigger social exclusion in online SE communities. Using comments from 189 projects that use Gitter as the communication platform, the work demonstrates the presence of all the 11 socially-exclusionary attributes. The paper also presents an outline of a framework called DARE that can help reduce such exclusionary messages in online communities and make them more welcoming for diverse users.

\bibliographystyle{ACM-Reference-Format}
\bibliography{sample-base}


\end{document}